\documentclass[12pt]{article}
\input{epsf}
\usepackage{amsmath,amssymb,graphicx}

\begin{document}

\newcommand{\RR}{\mathrm{I\!R\!}}
\newcommand{\II}{\mathrm{I\!I\!}}
\newcommand{\PP}{\mathrm{I\!P\!}}
\newcommand{\sn}{\mbox{sn}}
\newcommand{\am}{\mbox{am}}
\newcommand{\DEF}{\stackrel{\mbox{\rm\scriptsize def}}{=}}
\newcommand{\bm}[1]{\mbox{\boldmath $ #1 $}}

%\hyphenation{}

\def\R{\hbox{\rm I\kern-.2150em R}}
\def\B{\hbox{\rm I\kern-.2150em B}}
\def\maps{:}
\def\into{\rightarrow}
\def\QED{\vrule height 6 pt width 5 pt depth 0pt}

%Centre for Nonlinear Dynamics and its Applications,\\ 
%University College London, Gower Street,
%London WC1E 6BT, U.K.

%\title{Single DNA molecule experiments : a supercoiling model}
%\title{Single DNA molecule experiments : a contact model for supercoiling}
%\title{Getting DNA torsional rigidity from single molecule experiments}
\title{Getting DNA twist rigidity from single molecule experiments}
\author{
S\'ebastien Neukirch%\footnote{corresponding author,
%sebastien.neukirch@epfl.ch,
%tel~:~+41 21 693 27 66,
%fax~:~+41 21 693 55 30} 
\\
Bernoulli Mathematics Institute \\ % School of Basic Sciences, \\
Swiss Federal Institute of Technology, \\
%\'Ecole Polytechnique F\'ed\'erale de Lausanne,\\
CH-1015 Lausanne, Switzerland,\\
and\\
Laboratoire de Physique Statistique,\\
Ecole Normale Sup\'erieure, \\
F-75231 Paris cedex 05, France.%\footnote{present address}.
%\and \\
%Michael E. Henderson\footnote{mhender@us.ibm.com} \\
%T. J. Watson Research Center I.B.M., \\
%Yorktown Heights, NY 10598, U.S.A.
%
\date{\today}
%
%\date{}
%
}
%% %
%% \twocolumn[
%% %
%% \begin{@twocolumnfalse}
%% %
%
\maketitle
%
%
%\begin{center}
%\begin{tabular}{ll}
%s.neukirch@ucl.ac.uk & mhender@us.ibm.com \\ Centre for Nonlinear Dynamics &
%Mathematical Sciences\\ University College London & TJ Watson Research Centre\\
%London WC1 6BT, U.K. & U.S.A.\\
%\end{tabular}
%\end{center}
%
%
%
\begin{abstract}
We use an elastic rod model with contact to study
the extension versus rotation diagrams of single supercoiled
DNA molecules.
We reproduce quantitatively the supercoiling response of overtwisted DNA
and, using experimental data,
%
%This enables us to fit experimental data and to
%
%Hence by fitting experimental data with our model, 
we get an estimation of the effective supercoiling radius and of the twist
rigidity of B-DNA.
We find that unlike the bending rigidity, the twist rigidity
of DNA seems to vary widely with the nature and concentration of
the salt buffer in which it is immerged.
%
%% %
%% Fitting experimental extension-rotation diagrams of single molecule
%% experiments with an elastic rod model with contact enable us to get an
%% estimation of the effective supercoiling radius and of the stiffnesses ration
%% for DNA.
%% %
\end{abstract}
%
%
%\newpage

%% %
%% \end{@twocolumnfalse}
%% ]
%% %
%

%

%
%
% --- INTRODUCTION ---- %
%
%
At first the DNA molecule can be seen as the passive carrier of our genetic
code.
But  in order to understand how a $2 \;m$ long string of DNA can fit
into a $10 \; \mu m$ nucleus, it is clear that one has to
consider the mechanical properties of the molecule.
Namely the fact that the DNA double helix is a long and thin elastic filament
that can wrap around itself or other structures.
%
% chromatine, nucleosome, chromosomes.
%
Furthermore, the elastic properties of DNA play an important role in the
structural dynamics of cellular process such as replication and transcription.
Although the basic features of DNA were elucidated in the years following the
discovery of the double helix geometry, it is only during the past decade
that few groups, using different micro-techniques, have been able to manipulate
single DNA molecules in order to test their mechanical properties.
%

% --- EXPERIMENTS  ---%
%
A first way to manipulate a single molecule of DNA is to attach
a polystyrene bead at each of its two ends.
One bead is then stuck to
a micropipette and the other is held in an optical trap.
One pulls on the molecule by moving the pipette and measures the
force through the displacement in the trap.
%which acts like a spring.
%
This artefact is called an optical tweezer \cite{Quake97}.
Of course elastic properties of a DNA molecule will in general depend
on the sequence of base pairs (bp) it is made of.
Nevertheless for long molecules, i.e. more than a hundred bp,
the behavior of the molecule can be described by a coarse-grained model
known as the worm like chain \cite{kratky+porod:1949}.
In this model, DNA is considered as a semi-flexible polymer with bending
%flexural
persistance length $A$. This is the contour length over which correlations
between the orientation of two polymer segments is lost. It can be viewed as
the ratio of the elastic bending rigidity $K_0$ to the thermal energy $k_B \,
T$.
The common accepted value is $A=50 \; nm$ in physiological buffer.

\begin{figure}[htbp]
\begin{center}
\includegraphics[angle=0,width=0.65\linewidth]{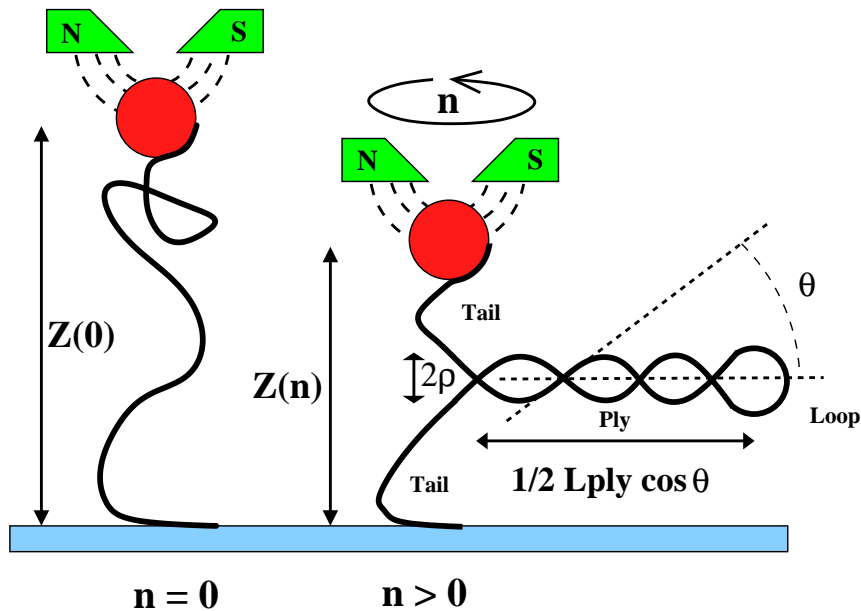}
\caption{The magnetic tweezer experiment.}
\label{fig magnetic tweezer}
\end{center}
\end{figure}

Another way of manipulating a single DNA molecule is to use
a magnetic tweezer \cite{strick+al:1996}.
Here the molecule is locked on a glass surface at one end, and
glued to a magnetic bead at the other end.
The bead is now controlled by a magnet that one rotates in order
to input a twist constraint in the system. 
The pulling force is tuned via the monitored distance between the magnet
and the bead. The end-to-end distance of the DNA molecule is
measured thanks to a microscope.
Experiments are carried under constant pulling force $f$. One gradually
rotates the magnet around an axis perpendicular to the glass surface while
monitoring the relative extension of the molecule $z=Z/L$ where $L$ is the
total contour length of the molecule and $Z$ is the distance in between the
bead and the glass surface, see fig. (\ref{fig magnetic tweezer}).
Force is measured using the Brownian motion of the bead.
No direct measurement of the twist moment is possible
with magnetic tweezers. Consequently the twist persistence length
$C$ is not directly available.

When no rotation is put in, the DNA molecule behaves like a semi-flexible
polymer, i.e. the relative extension $z$ is a function of the temperature $T$,
the persistence length $A$ and the applied pulling force $f$:
\begin{equation}
z(n=0)=1-\left( \frac{4 \, k_B \, T}{A\, f}\right)^{\frac{1}{2}}.
\label{equa z wlc}
\end{equation}
This relation is used to extract the bending persistence length $A$ from
experimental data.
Then under gradually increased rotation, the extension $z$
decreases with the number of turns $n$ put in and eventually the molecule
starts to wrap around itself.
Geometrically speaking, the DNA molecule is coiling around itself
in a helical way. Since the molecule is already a double helix, we
speak about supercoiling.
Each helical wave of the super helix is called a plectoneme.
Different attempts has been made to model this experiment, introducing the
concepts of wormlike rod chain \cite{bouchiat+mezard:2000}, or torsional
directed random walk \cite{moroz+nelson:1997} but without taking into account
the possibility of contact, or dealing with the plectonemes in an ad hoc way
\cite{marko+siggia:1995}.
%

%
%
% --- MODEL ---- %
%
%
Here we present an elastic model that specifically includes the self-contact
of the DNA molecule, but does not take into account thermal fluctuations. Our
point is that, in the regime where plectonemes are formed, the relevant
physical information is already present in our zero-temperature elastic rod
model with hard-wall contact.
Due to the base pair conformation, it seems natural to consider DNA 
as an elastic rod with a non-symmetric cross-section (i.e. two different
bending rigidities).
Nevertheless it has been shown that when dealing with a long enough molecule
(more than a few dozen base pairs), one could simplify the model and work with
an effective rod having a symmetric cross-section (i.e. one bending rigidity
that is the harmonic mean of the two raw rigidities)
\cite{kehrbaum+maddocks:2000}.
%\cite{rey:2002}.
%
In order to focus on supercoiling, we consider the simplest elastic rod model
than includes twist effects and can have 3D shapes.
%
%
% --- Elastic Kirchhoff rod with self-contact and boundary conditions --- %
%
%
Following the classic terminology of {\it Euler planar elastica} for twistless
2D shapes, we call the present model the {\it Kirhhoff ideal elastica}
\cite{neukirch+henderson:2002}. The elastic energy reads:
\begin{equation}
E = \frac{1}{2} \, \int_0^L \left( K_0 \, \kappa^2(s) + K_3 \, \tau^2(s)
\right) ds \nonumber
\end{equation}
where $s$ is the arc-length, $\kappa(s)$ the curvature of the centre line,
$\tau(s)$ the twist rate of the cross section around the centre line (which in
the case of an ideal elastica is a constant of $s$), and $K_0$ and $K_3$ are
the bending and twist rigidities respectively.
The Kirchhoff equilibrium equations read:
\begin{eqnarray}
\bm{F}'(s)+\bm{p}(s) & = & \bm{0} \label{equa force balance}\\
\bm{M}'(s)+\bm{R}'(s) \times \bm{F}(s) & = & \bm{0}
\end{eqnarray}
where $\bm{F}(s)$ and $\bm{M}(s)$ are the internal force and moment
respectively. 
The external force per unit length $\bm{p}(s)$ can model an
electrostatic repulsion, gravity or hard-wall contact.
The centre line of the rod is given by $\bm{R}(s)$ and $\bm{t}(s)=\bm{R}'(s)$
is its tangent.
In the case of an ideal rod it can be shown \cite{heijden+al:2001} that:
\begin{eqnarray}
K_0 \, \bm{t}'(s) & = & \bm{M}(s) \times \bm{t}(s)\\
K_0 \, \bm{d_1}'(s) & = & \left( \bm{M}(s) - \tau \left( K_3 - K_0 \right)
\bm{t}(s)\right) \times \bm{d_1}(s)
\end{eqnarray}
where $\bm{d_1}(s)$ is a unit vector, lying in the cross section, that
%(i.e. perpendicular to $\bm{t}$)
follows the twist of the centre line. For a DNA molecule,
it is generally taken as the vector pointing toward the major groove.
\begin{figure}[htbp]
\begin{center}
\includegraphics[angle=0,width=0.50 \linewidth]{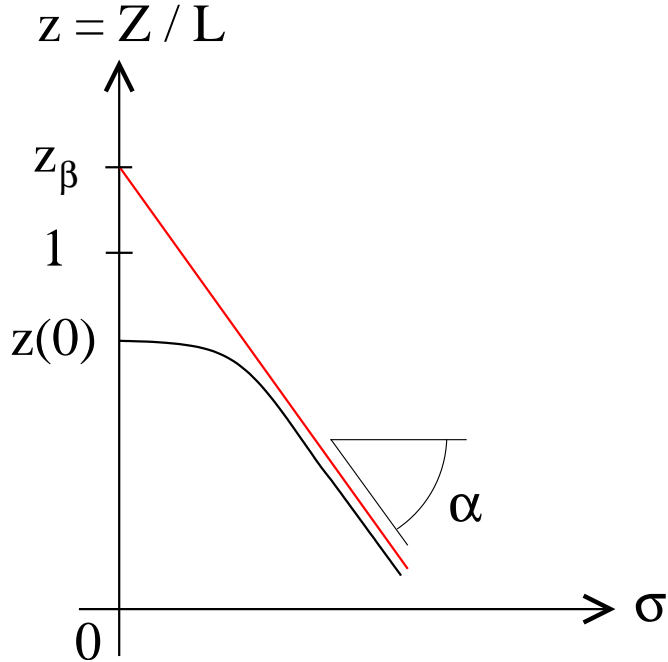}
\caption{Fitting the linear part of the response curve.}
\label{fig plecto}
\end{center}
\end{figure}
For the parts free of contact, we have $\bm{p}(s) \equiv \bm{0}$.
In case of self-contact there are two points along the rod, say at arclength
$s_1$ and $s_2$, where the inter-strand distance $|\bm{R}(s_1)-\bm{R}(s_2)|$
is equal to twice the radius of the circular cross-section (which we denote by
$\rho$).
At point $s_1$, we introduce a finite jump in the force vector
$\bm{F}(s)$:
\begin{equation}
\bm{F}(s<s_1) = \bm{F}(s>s_1) + \frac{ \Delta F_{12} }{2 \rho} \, \left(
\bm{R}(s_1) - \bm{R}(s_2) \right)
\end{equation}
where $\Delta F_{12}$ is a positive real number.
The same treatment is done at point $s_2$, with
the same $\Delta F_{12}$ \cite{heijden+al:2001,coleman+swigon:2000}.
This corresponds to having a Dirac function for $\bm{p}(s)$ in (\ref{equa
force balance}). In case of continuous sections of contact $\bm{p}(s)$
is a function with changing direction and intensity.
In our model we only consider cases where the contact occurs either at 
discrete points or along straight lines.
This model has already been used \cite{stump+al:1998,gaspar+nemeth:2001}
and is well described in \cite{coleman+swigon:2000}.
%

%
%
% --- Results of continuation --- %
%
%
Ideally we are looking for equilibrium configurations of the ideal elastica
that match the boundary conditions corresponding to the magnetic tweezer
experiment.
Nevertheless, in case of a long rod ($L \gg \rho$) with a large number of
turns $n$ put in, the exact geometry of the boundary conditions are less
important, and we choose clamped ends as used before \cite{heijden+al:2001}.
We numerically find equilibrium configurations matching the boundary
conditions using classical path following techniques; first a self made
algorithm relying on multiple (or parallel) shooting, %\cite{keller:1976},
then using the code AUTO \cite{Doedel:1991} that discretises the boundary
value problem with Gauss collocation points.
%orthogonal collocation at Gaussian points.
%
% --- Numeric continuation : main parameters : $(sr,T)$ --- %
%
The different types of solutions (straight, buckled, supercoiled) are found
in the following way. We fix the radius $\rho$ and the vertical component $f$
of the force vector $\bm{F}(s=0)$ acting on the bead.
We first consider a straight rod with no rotation $n=0$. 
Then we twist the rod by gradually increasing $n$. At a certain
bifurcation value, the path of straight solutions crosses another path of
buckled solutions.
Following this new path, we end up crossing another path of solutions
with one discrete contact point.
This latter path will itself intersect a path of configurations with two
contact points.
Solutions with up to three discrete contact points have been found
\cite{heijden+al:2001}.
Eventually the configurations with three discrete contact points
bifurcate to solutions including a line of contact in addition of discrete
contact points.
%\cite{coleman+swigon:2000}
We call them {\it supercoiled configurations}.
In a supercoiled configuration, the parts of the rod which are in
continuous contact have an helical shape.
We call this twin super-helix a {\it ply}. 
The super-helix is defined by its radius $\rho$ and its helical angle $\theta$
(see fig. (\ref{fig magnetic tweezer})).
Each time we change the fixed pulling force $f$ or the radius $\rho$, the
entire numerical continuation has to be rerun.
Since we do not consider thermal fluctuations, the first part of our numerical
response curve does not correspond to what is found experimentally.
On the other hand our model reproduces quite precisely the part of the
response curve where the distance $z$ decreases linearly with the number of
turns $n$, provided we identify $\rho$ not with the crystallographic radius of
the DNA molecule but with an effective supercoiling radius due to
electrostatic as well as entropic repulsion.
%
%
%
% --- Saturation at high $\sigma$ --- %
%
%
% --- non constant $\theta(s)$ angle --- %
%
%
%
%
%
% --- FIT ---- %
%
We numerically find that in the linear regime the helical angle $\theta$ does
not vary with $n$.
We fit numerical solutions curves as in \cite{stump+al:2000} and we find that
$\theta$ only depends on $f$, $K_0$, and $\rho$:
\begin{equation}
f=\frac{K_0}{\rho^2} \, \phi_3(\theta) \mbox{ with } 
\phi_3(\theta)=1.65805 \, \theta^4 \label{equa T R theta} \, .
\end{equation}
%
%

%--------------------------------------------------------------------%

This result enables us to extract the effective supercoiling radius $\rho$
and the twist rigidity $K_3$ from magnetic tweezer experiments on DNA.
First we note that the number of turns $n$ applied to the magnetic bead can be
interpreted as the excess link of the DNA molecule: $n=\Delta Lk$.
Link is normally defined for a closed ribbon but carefull use of a closure
permits the introduction of the link of an open DNA molecule
\cite{starostin:2003,rossetto+maggs:2003}.
We use the C\u{a}lug\u{a}reanu-White-Fuller theorem to decompose the excess link:
\begin{equation}
(n=) \, \Delta Lk=\Delta Tw + Wr \label{eq calugareanu}
\end{equation}
where $\Delta Lk$ (resp. $\Delta Tw$) is the difference between
the actual link (resp. twist) and the intrinsic link (resp. twist) of the
double helix.
The writhe $Wr$ is the average number of crossings of the centre line one sees
when looking at the molecule from (all the possible) different viewpoints.
\begin{figure}[htbp] 
$$ \epsfxsize=0.55\linewidth \epsfbox{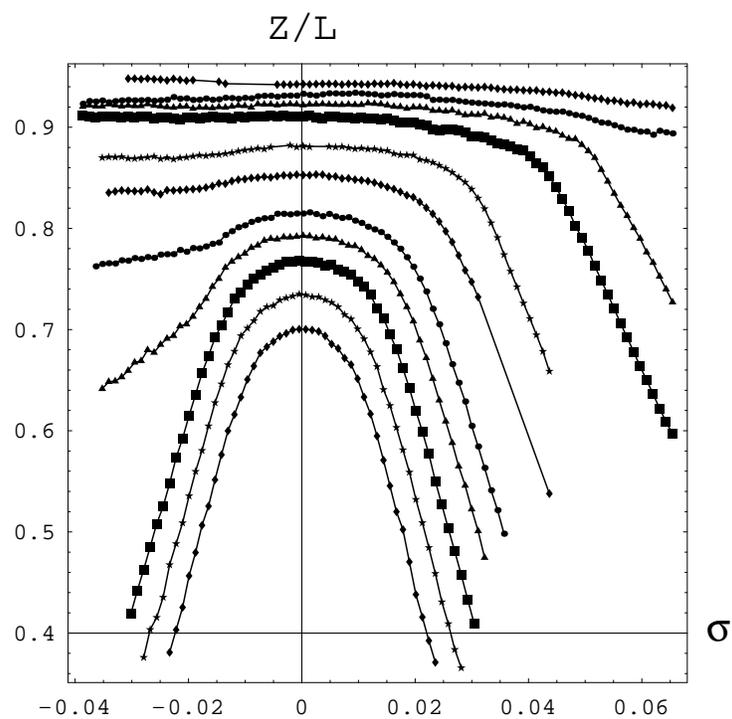} $$
\caption{Experimental response curves with a 48kbp DNA molecule in a 10 mM
phosphate buffer \cite{strick+al:1996}.  Each curve corresponds to an
experiment carried at a constant pulling force. Forces are, from bottom to
top, 0.25, 0.33, 0.44, 0.57, 0.74, 1.1, 1.31, 2.2, 2.95, 4.15, and 4.9 pN.}
\label{fig 48kbp}
\end{figure}
In the ply part, the twist rate is related to the helical angle $\theta$
via a mechanical balance \cite{coleman+swigon:2000,neukirch+heijden:2002}:
\begin{equation}
%\Delta Tw 
\tau = (-\epsilon) \frac{K_0}{2\rho \, K_3} \,
\left( \tan 2\theta-\sin 2\theta \right) \, ,
\label{eq twist ply}
\end{equation}
where $\epsilon=\pm 1$ stands for the chirality of the ply
\cite{neukirch+heijden:2002}.
Since the twist rate is constant along the rod, we have $\Delta Tw=\tau \,
L/(2\pi)$.
Generally writhe is not additive but using Fuller theorem \cite{aldinger:1994}
with a carefully choosen reference curve we may write :
\begin{equation}
Wr=Wr_{loop}+Wr_{tails}+Wr_{ply}.
\end{equation}
Here we neglect $Wr_{loop}$ and $Wr_{tails}$.
Carefully choosing a reference curve to apply Fuller theorem or 
directly computing the writhe from the double integral yields 
\cite{starostin:2003}:
\begin{equation}
Wr_{ply}=(-\epsilon) \frac{L_{ply}}{4\pi \,\rho} \sin 2\theta \, .
\label{eq writhe ply}
\end{equation}
The total contour length (or number of base pairs) $L$ of the DNA molecule
is given and we write:
\begin{equation}
L_{ply}=L-L_{loop}-L_{tails} \, .
\end{equation}
We neglect $L_{loop}$ and we use an heuristic equation for
$ L_{tails} $ :
\begin{equation}
L_{tails}= Z(\sigma) \, \frac{L}{Z(0)},
\end{equation}
which means that when the molecule is supercoiled the tails parts, which are
not supercoiled, are as disordered as the molecule was when no rotation was put
in.
For positive supercoiling ($n>0$) we get a left handed ply ($\epsilon=-1$).
From (\ref{eq calugareanu}), (\ref{eq twist ply}), and (\ref{eq writhe ply})
we obtain:
\begin{equation}
\Delta Lk=
\frac{L}{4\pi \rho} \left[ 
\frac{K_0}{K_3} \, \left( \tan 2\theta-\sin 2\theta \right)+ 
\sin 2 \theta \left(1-\frac{Z}{Z(0)}\right)
\right].
\label{eq link total}
\end{equation}
Introducing the supercoiling ratio $\sigma=\Delta Lk/Lk_0$ and
taking $h=10.5$ base pairs per turn and 
$\delta z=0.34 \, nm$ of rise for each base pair, we get an
helical pitch for the DNA double helix $H=h \, \delta z = 3.57 \, nm$ and 
we have: $Lk_0=L/H$.
%$\Delta Lk= \sigma L / H$.
%
Equation (\ref{eq link total}) can then be inverted to give an approximation
of the linear part of the response curve in the $(\sigma,z)$ plane :
\begin{equation}
\frac{z}{z(0)}=
%\frac{Z}{Z(0)}=
1+
\frac{K_0}{K_3}\left( \frac{1}{\cos 2 \theta}-1\right)-
\frac{4 \pi \rho}{H \sin 2 \theta} \, \sigma \label{equa linear part}
\end{equation}
\begin{table}[htbp]
\begin{center}
\begin{displaymath}
\begin{array}{|c|c|c|c|}
\hline
f \mbox{ (pN)} & \theta \mbox{ (rad)} & \rho \mbox{(nm)} & K_3/K_0 \\
\hline
0.25   &   0.427    &    6.85  &   1.88 \\
0.33   &   0.449    &    6.60  &   1.87 \\
0.44   &   0.467    &    6.17  &   1.92 \\
0.57   &   0.469    &    5.47  &   1.84 \\
0.74   &   0.504    &    5.55  &   2.01 \\
1.1    &   0.488    &    4.26  &   1.86 \\
1.31   &   0.471    &    3.64  &   1.58 \\
2.2    &   0.503    &    3.21  &   1.65 \\
2.95   &   0.507    &    2.81  &   1.62 \\
\hline
\end{array}
\end{displaymath}
\end{center}
\caption{Results for the 48kbp DNA molecule in a 10 mM phosphate buffer.}
\label{table monovalent}
\end{table}
\begin{figure}[htbp] 
$$ \epsfxsize=0.55\linewidth \epsfbox{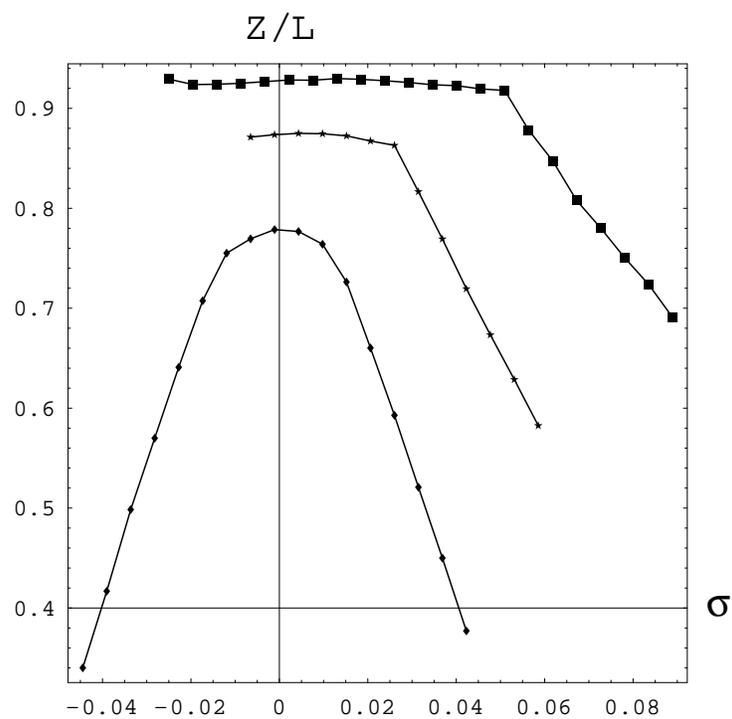} $$
\caption{Experimental response curves with a 11kbp DNA molecule in a 150 mM
phosphate and 5 mM magnesium (Mg2+) buffer.  Forces are, from bottom to top,
0.45, 1.45, and 4.3 pN. Unpublished data from Gilles Charvin.}
\label{fig 11kbp}
\end{figure}
%
%
%
%
% --- RESULTS ---- %
%
%
We use (\ref{equa T R theta}) and (\ref{equa linear part}) to extract
information from experimental response curves of extension-rotation
experiments.
Since experiments are carried at given (fixed) $f$ and  
since $K_0$ is obtained from (\ref{equa z wlc}) with $K_0=A \, k_B T$,
then from each experimental curve in the $(\sigma,z)$ plane we only need : 
$(i)$ the relative extension at $\sigma=0$, which we denote by $z(0)$,
$(ii)$ the slope of the linear part, which we denote by $\alpha$,
$(iii)$ the ordinate at the origin of the straight line fitting
the linear part of the response curve, which we denote by $z_{\beta}$.
Using (\ref{equa T R theta}) and (\ref{equa linear part})
%and the experimental value for $\alpha$,
we get an equation for $\theta$ :
\begin{equation} 
\alpha=-\frac{4\pi \, z(0)}{H \sin 2 \theta} \, 
\sqrt{\frac{K_0}{f} \phi_3(\theta)} 
\end{equation}
Once we know $\theta$, 
we can extract the effective supercoiling radius
$\rho$ from (\ref{equa T R theta}) 
and the effective stifnesses ratio from (\ref{equa linear part}):
\begin{equation} 
\frac{K_3}{K_0} = \left(\frac{1}{\cos 2\theta}-1 \right) \Big/
\left(\frac{z_{\beta}}{z(0)}-1 \right)
\end{equation}
\begin{table}[htbp]
\begin{center}
\begin{displaymath}
\begin{array}{|c|c|c|c|}
\hline
f  \mbox{ (pN)} & \theta \mbox{ (rad)} & \rho \mbox{ (nm)} & K_3/K_0 \\
\hline
0.45  &   0.307  &  2.79  &   1.09 \\
1.45  &   0.319  &  1.67  &   1.00 \\
4.3   &   0.348  &  1.15  &   0.99 \\
\hline
\end{array}
\end{displaymath}
\end{center}
\caption{Results for the 11kbp DNA molecule in 
a 150 mM phosphate and 5 mM magnesium (Mg2+) buffer.}
\label{table bivalent}
\end{table}
%
%
%
%
%

%
%
%
% --- Conclusion --- %
%
%
%
Results are shown in table \ref{table monovalent} and \ref{table bivalent}
and can be compared to results of \cite{bryant+al:2003} where another
micro-technique was used and a value of $K_3 \simeq 100 \, nm \, k_B \, T$ (in a
100mM NaCl + 40 mM Tris-HCl buffer) was found.
From our results, we see that the effective supercoiling radius decreases with
the intensity of the pulling force, and with the strength of the salt buffer.
Fitting experimental data with (\ref{equa z wlc}) yields $K_0 = 51 \, nm \,
k_B \, T$ for fig. \ref{fig 48kbp} and $K_0 = 57 \, nm \, k_B \, T$ for
fig. \ref{fig 11kbp}.
This means that the bending rigidity does not vary widely with the salt
properties of the buffer.
On the other hand, present results imply that the twist rigidity
strongly decreases with the salt strength.
This in turns implies that the electrostatic contribution (repulsion of the
charges lying on the sugar-phosphate backbone) to this effective twist
rigidity is rather important.

It is a pleasure to thank G. Charvin, V. Croquette, and D. Bensimon for
supplying us with data from experiments.

\end{document}